\documentclass{article}

\usepackage{arxiv}

\usepackage[utf8]{inputenc} 
\usepackage[T1]{fontenc}    
\usepackage{hyperref}       
\usepackage{url}            
\usepackage{booktabs}       
\usepackage{amsfonts}       
\usepackage{nicefrac}       
\usepackage{microtype}      
\usepackage{cleveref}       
\usepackage{graphicx}
\usepackage{xcolor, soul} 
\usepackage[title]{appendix}
\usepackage[backend=bibtex,style=ieee,style=numeric-comp, sorting=none, maxnames=5]{biblatex} 
\graphicspath{{./}{./figures/}} 

\title{Impossible Cloud Network: A decentralized \\ Internet Infrastructure Layer}

\date{October 1, 2025}

\newif\ifuniqueAffiliation
\uniqueAffiliationtrue

\ifuniqueAffiliation 
\author{Siu Kei Chung, Dr. Francisco Carpio, Andrei Navoichyk, Siarhei Valasovich,  \\ 
Dr. Jordan Moore, Dr. Slobodan Sudaric-Hefner, Daniel Baker, \\ 
Dr. Thomas Demoor, Maurizio Binello, Dr. Christian Kaul and Dr. Kai Wawrzinek
    \And
	Impossible Cloud Network  
}
\else
\fi

\renewcommand{\headeright}{Whitepaper v1.1}
\renewcommand{\undertitle}{Whitepaper  v1.1}
\renewcommand{\shorttitle}{ICN: A decentralized Internet Infrastructure Layer}

\begin{document}
\maketitle

\begin{abstract}
	The internet faces a sovereignty crisis due to power concentration and data
	growth among a few hyperscalers, leading to centralization and loss of user
	control. This consolidation risks censorship and creates single points of
	failure. While Web3 offers decentralized solutions, they often sacrifice
	either scalability, decentralization, or security, which are key elements in
	the blockchain trilemma. These solutions also struggle with limited access
	to enterprise-grade hardware and frequently rely on centralized
	infrastructure. The Impossible Cloud Network (ICN) addresses these issues by
	creating a multi-tiered, decentralized infrastructure layer. ICN offers a
	composable service layer, an enterprise-grade hardware resource layer, and a
	transparent, permissionless HyperNode network for performance enforcement.
	By strategically decoupling and decentralizing each layer, ICN aims to
	provide an open, extensively scalable infrastructure that ensures digital
	sovereignty, eliminates single points of trust, enables service
	programmability, and offers a decoupled architecture for limitless
	possibilities in the future internet.
\end{abstract}


\section{Introduction}

Concentration of power and data growth among a few hyperscalers is leading to
centralization, single points of trust, and user data control loss. Hyperscaler
infrastructure, while providing scaling benefits, also risks consolidating
control, raising censorship, access concerns, especially for personal data, and
creating single points of failure \cite{5283911}. Existing cloud providers
restrict innovation and resource utilization due to hardware and proprietary
software limitations, hindering the global cloud network's value and user
capabilities. This market is dominated by a few hyperscalers with proprietary,
verticalized technology stacks, resulting in vendor lock-in, interoperability
issues, and barriers to market entry \cite{8030657},  allowing them to dictate
pricing.

Emerging Web3 projects offer community-driven, open-source alternatives,
prioritizing transparency, affordability, and user-centric ecosystems with
respected data ownership. However, these solutions continue to suffer from
mainstream adoption problems and are unable to compete with existing Web2
alternatives owing to performance limitations inherent in many crypto-native
architectures \cite{10701160}. 

Web3 aims for decentralization via blockchain, but relies on concentrated
infrastructure, creating a single point of failure. Despite appearing
decentralized, many projects depend on a few cloud providers, risking widespread
application disruption if the infrastructure fails
\cite{10.1145/3618257.3624797}. While decentralization brings many benefits, the
lack of central political authorities makes the current Web3-based cloud
services miss the offering of Service Level Agreements (SLAs) that provide
customer guarantees \cite{10.1145/3656015}, which traditional business customer
models rely on. Traditional cloud service providers, on the other hand, provide
SLA guarantees but with self-measured key performance indicators, creating an
inherent lack of accountability, with no independent verification, leaving
customers to just rely on trust \cite{8737580,10.1145/2331576.2331586}. 

Impossible Cloud Network (ICN) overcomes these challenges through transparently
verifiable and enforceable hardware quality standards that allow the network to
compete with hyperscaler level performance and push beyond the current
boundaries of Web3. ICN envisions a multi-tiered network ecosystem, forging
essential building blocks for the next-generation of the internet’s evolution.
ICN features a highly adaptable application layer at the service level,
facilitating limitless service functionalities. This is achieved through a
hyperscaled resource layer at the hardware level, supported by diverse hardware
types. Bridging these layers is the HyperNode network, offering transparent and
permissionless security maintained by a broad community. By strategically
separating and decentralizing each layer, ICN unlocks a genuinely open and
extensively scalable infrastructure that empowers boundless capabilities for the
future of the internet. ICN focuses on four key areas:
\begin{itemize}
	\item Digital Sovereignty: Decentralizing the cloud through ICN enables
	digital sovereignty and mutual accountability by decoupling physical resources
	from digital services. This fosters flexibility, choice, and new applications
	like self-sovereign identity and an uncensorable internet, ultimately
	recapturing user trust and creating a more equitable, interconnected society. 
	\item Eliminating single-points-of-trust: The ICN protocol offers unlimited
	scalability, allowing the network to control its growth. It supports diverse
	hardware via defined classes, enabling efficient resource allocation at varying
	performance levels. This creates a unified global hardware layer for all types
	of applications. Ultimately, ICN aims to be an open world compute platform,
	efficiently allocating diverse hardware and providing a resource-aware base for
	the future internet. 
	\item Service programmability: By offering the resource layer as a
	programmable foundation, the ICN Protocol aims to grow a rich ecosystem of
	applications and services. This empowers entities to develop resource-aware
	applications, finely tuned for both the underlying hardware and the target
	users. This capability unlocks the potential for value creation to its fullest
	human extent, eliminating limitations on what can be constructed and provided on
	the internet of the future.
	\item  Decoupled architecture: ICN's decoupled architecture
	enables service composition by separating hardware, resource management,
	services, and applications. This allows independent software solutions to
	leverage ICN's API, dynamically compose resources, and offer plug-and-play
	service integration.
\end{itemize}

\section{The ICN Protocol}

The ICN Protocol is divided into five fundamental layers:
\begin{itemize}
	\item Hardware: The underlying physical infrastructure of the ICN supplied
	by Hardware Providers, which are responsible for managing the physical
	network and hardware resources.
	\item Resource Composition: A physical resource abstraction layer to
	decompose our hardware components into logically separate resource units.
	This recomposes resources into usable instances for software deployment.
	\item Performance Enforcement: A decentralized network of nodes that
	periodically monitor and report the performance of the provisioned hardware.
	This layer ensures data availability and on-chain proof. 
	\item Services: Standalone and integrated software solutions deployed by
	third parties on top of ICN resources.
	\item Applications: End use cases or other products that integrate with
	services deployed on ICN. These applications may use resources or
	communicate with APIs from ICN services.
\end{itemize}

\begin{figure}[h]
	\centering
	\includegraphics[width=0.85\columnwidth]{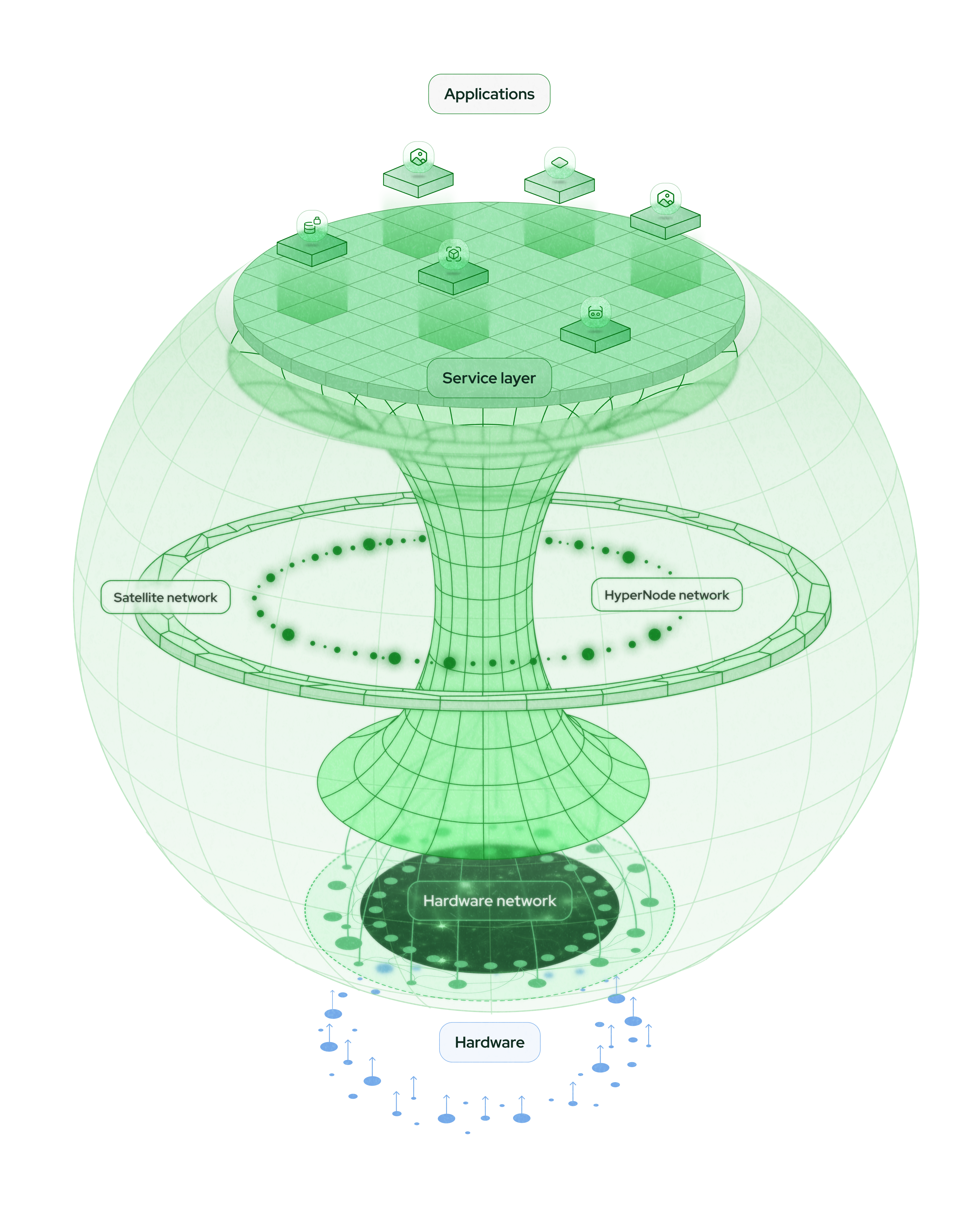}
	\caption{Layers in the ICN Protocol}
\end{figure}

The blockchain serves as the ultimate immutability layer for the ICN protocol,
managing resource allocation to prevent double spending of hardware and ensure
correct user assignment. It also governs staking, collateral, and slashing
mechanisms to enforce protocol rules and maintain system security.

\subsection{Hardware layer}

The Hardware Layer in the ICN protocol is a globally accessible, distributed
resource pool built upon ScalerNodes. These are the smallest physical units of
enterprise-grade hardware resources committed by hardware providers, who are
compensated for their contribution. Each ScalerNode can be of any hardware class
depending on the type of resource it commits to the network and are rewarded
accordingly to the provided capacity. Service Providers and Builders can combine
ScalerNodes into larger macronodes with integrated capabilities. This layer
forms the fundamental physical infrastructure required for the protocol's
operation.

\begin{figure}[h]
	\centering
	\includegraphics[width=1.0\columnwidth]{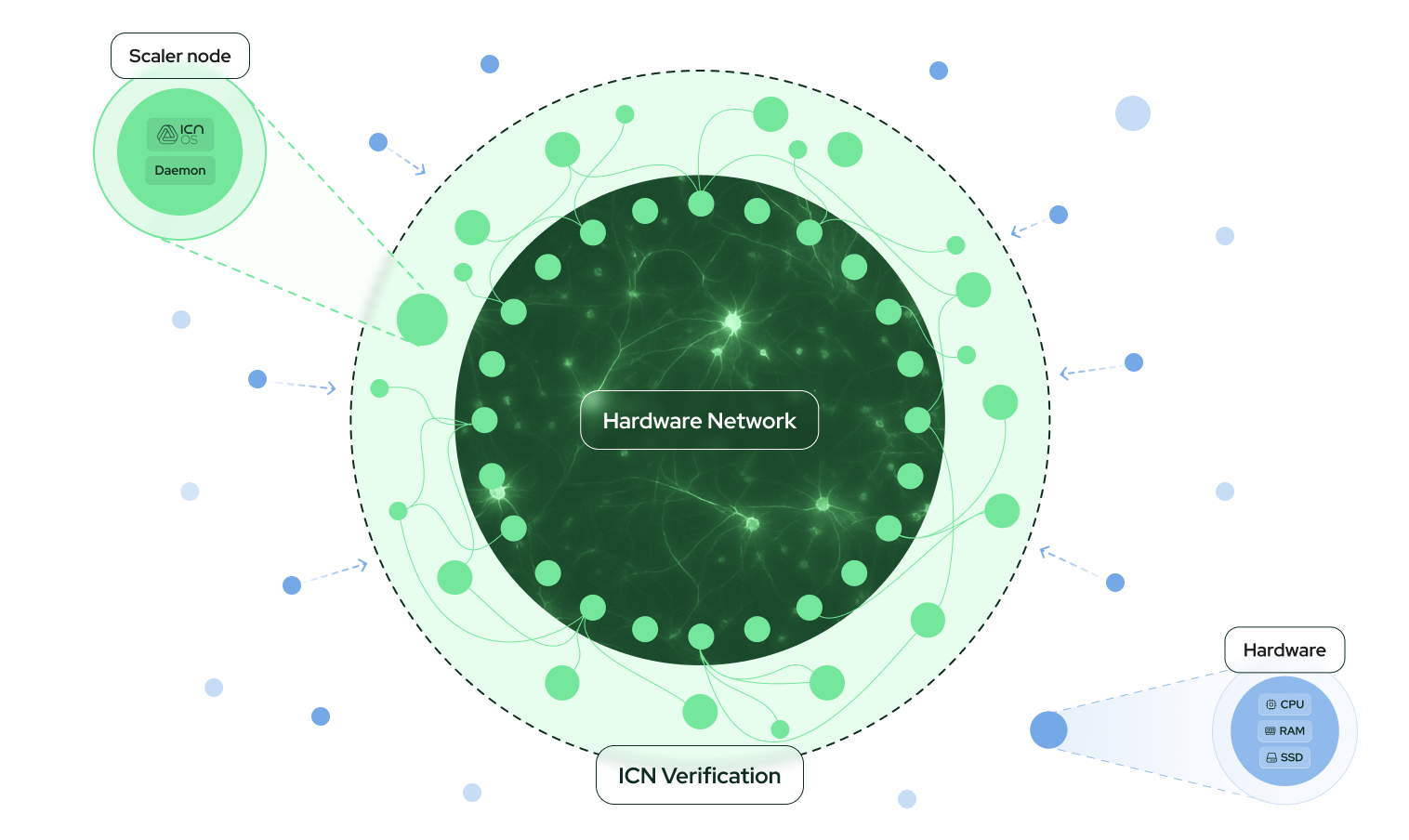}
	\caption{Hardware Layer}
\end{figure}

\subsubsection{ScalerNodes: Decentralized hardware resources}

Hardware providers are all operationally decentralized, and secured through
collateral. This way we can ensure that there are no central points of trust or
failure and allow an unparalleled level of composability and optionality that
can respond dynamically to market needs. Hardware providers commit nodes for
certain periods of time to guarantee that the network maintains certain
capabilities for known durations and enables long term uptime of services while
allowing flexibility for hardware providers to manage their own resource
commitments versus rewards.

ScalerNodes are contributed to ICN by Hardware Providers which manage the
underlying physical infrastructure, ensuring secure remote access. Their main
responsibilities include physical host maintenance (including firmware updates
and attached devices), automated OS bootstrapping, internal network
configuration (including routers, switches, firewalls, etc), and external
internet connectivity. Hardware Providers register ScalerNodes in ICN by
specifying the hardware class, location, capacity, rewards share, reservation
price, and maximum booking duration. The registration of ScalerNodes requires
collateralization by the Hardware Providers to ensure the long-term commitment
to the network.

\subsubsection{Global resource pool}

Our hardware network forms an interconnected mesh of ScalerNodes that are
globally distributed. This network acts as a base layer of infrastructure
resources on which services and applications can be run. ICN has devised the
hardware network to be a ubiquitous, “always-on” constant that users can always
contact and communicate with.

The representation of the global resource pool is as a localized capability map
of hardware and allows these resources to be used and allocated on an efficient
basis by exposing a “resource-aware” abstraction to the software layer. This
enables us to construct highly performant and efficient, fit-for-purpose
instances for application deployment by the Resource Composition Layer.

\subsection{Resource Composition layer}

The Resource Composition Layer is a dynamic hardware allocation layer that
utilizes an abstraction engine to decouple resource capabilities from the
underlying hardware component. While ScalerNodes provide physical units of
resources in the form of disks, processors and memory, the Resource Composition
Layer creates logical units of resources so that all available capacity is
allocated efficiently.

ScalerNodes are decomposed into fundamental resource units defined by type:
storage, compute, memory, networking. These units contribute to a globally
distributed pool that is accessible permissionlessly. Each resource pool in our
system represents a unique capability provided by our ScalerNodes and may each
also contain further designation depending on the type e.g. fast or slow
storage.

\begin{figure}[h]
	\centering
	\includegraphics[width=0.7\columnwidth]{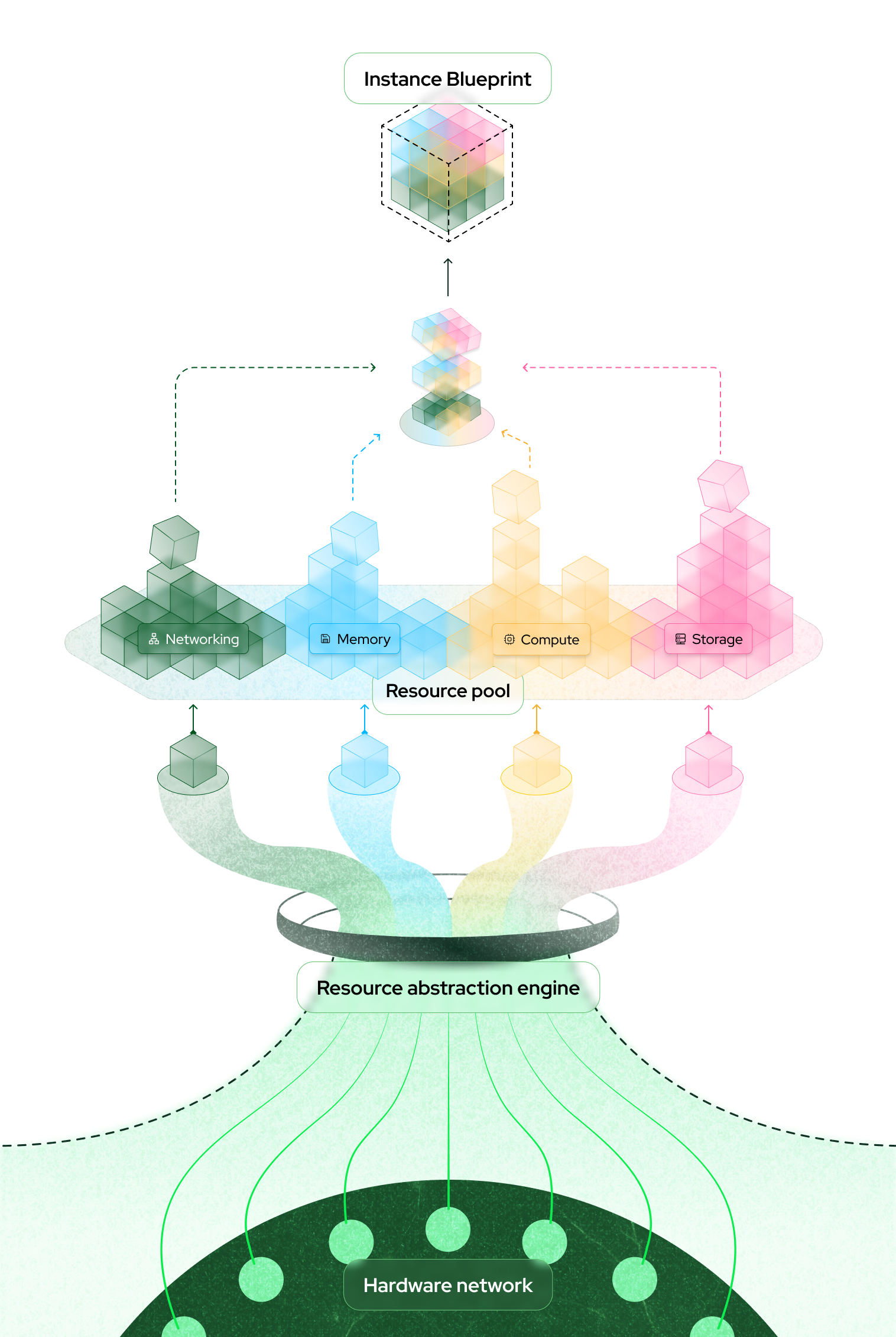}
	\caption{Resource Composition Layer}
\end{figure}

\subsubsection{Resource composability}

Hardware capability provided by ScalerNodes are pooled into tangible resources
by the global resource pool. The Resource Abstraction Engine arbitrarily
recomposes available resource units into instances defined by Instance
Blueprints which describe compositional configurations for available resources
by type. Blueprints are used to construct and spin-up usable execution engines
for application deployment and operation.

Instance blueprints are created as commonly used configurations based on the use
cases or hardware setup of the ScalerNode to improve efficiency. Some instance
configurations may be more or less performant based on a trilemma: use case
needs, available resources, physical configuration.

\subsubsection{Elastic Instances}

Resource units can be composed in any configuration, constrained by the locality
of the resource depending on the requirements specified. For example, if a
high-storage instance is required and it must only store data in Europe, it is
impossible to compose storage resource units from outside of Europe. Performance
requirements may also provide similar constraints to defining instance
configurations.

Instances can also be configured to expand and contract subject to available
resources. Instances typically will operate from a static configuration, but can
be defined to allow to flex in response to usage as long as the resources are
not allocated elsewhere or otherwise unavailable. Each resource type will follow
different rules for dynamic scaling.

\subsubsection{On Demand Deploy}

Users can specify a Blueprint or configure a custom configuration and request a
deployment from ICN. The Resource Composition Layer will accept the
specifications provided by the user and construct an instance with the required
storage, compute, memory and networking resources. Instance deployments are
subject to requirement and availability constraints.

\subsection{Performance Enforcement layer}

ICN employs a novel approach to performance assurance through a decentralized
validator network known as the HyperNode Network. Data gathered by the
HyperNodes is temporarily accessible and auditable on a data availability
network called the Satellite Network. HyperNodes generate performance
assessments of the hardware layer and subsequently submit cryptographic proofs
to the blockchain. This innovative approach aims to establish a more robust and
dependable framework for ensuring consistent performance within decentralized
environments.

\begin{figure}[h]
	\centering
	\includegraphics[width=0.7\columnwidth]{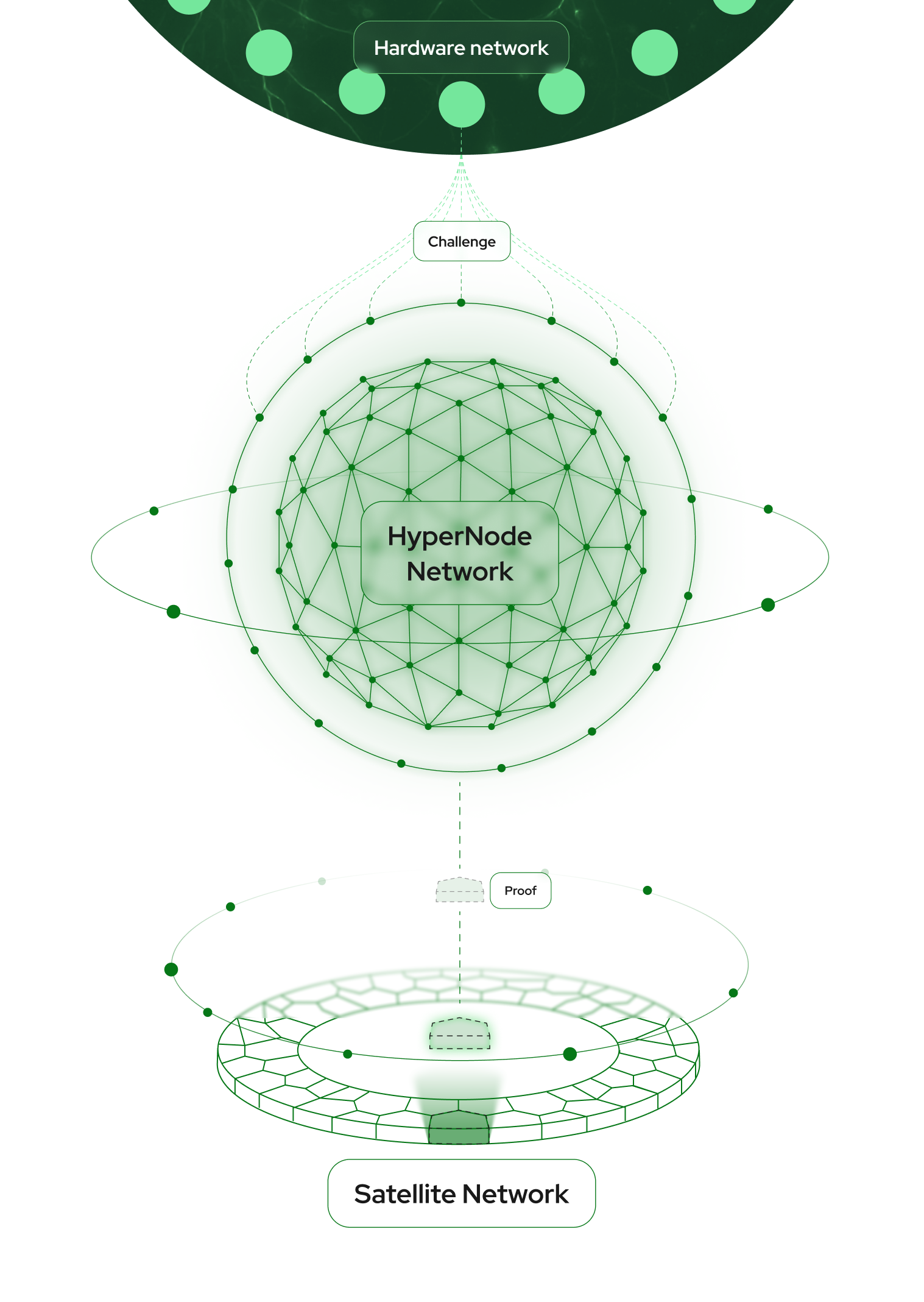}
	\caption{Performance Enforcement Layer}
\end{figure}

\subsubsection{Decentralized Monitoring Network}

The HyperNode network is crucial for maintaining the performance and security of
the infrastructure. Each HyperNode acts as a third-party entity which
independently checks and verifies the performance of every ScalerNode in the
hardware layer, guaranteeing the reliable and consistent operation of services
and applications. 

HyperNodes execute different challenges according to each type of hardware class
of the ScalerNode to verify specific performance capabilities of the node. Each
challenge generates a set of reports with key performance indicators that are
then published regularly to the Satellite Network, and proofs are submitted
on-chain for data attestation. 

\subsubsection{Data Availability Network}

The Satellite Network is a key component in maintaining the auditability and
transparency within the ICN. Allowing public verifiability is a necessary
requirement for a truly decentralized system, giving users the agency to monitor
and validate information in a trustless way.

The Satellite Network in ICN enables data availability for the HyperNode reports
to maintain transparent access to proof data required to validate the HyperNode
report results. This allows any user to confirm that the reports committed to
the Satellite Network align with the cryptographic proofs submitted on the
blockchain, fostering trust and accountability throughout the network.

\subsubsection{Pluggable proofs}

Proofs within the ICN Protocol transcend the hardware layer, extending their
applicability and functionality to services. This approach allows for the
implementation of highly specific proofs tailored to individual services,
enabling a flexible validation ecosystem. The HyperNode network provides a
standardized interface that facilitates the seamless execution of these
service-level proofs. This innovative design ensures that the integrity and
authenticity of operations are verifiable not only at the foundational hardware
level but also throughout the diverse range of services offered within the ICN
network.

\subsection{Service layer}

Service providers and builders can leverage ICN's abstraction layer to deploy
services on its decentralized hardware. They access instances by booking
customizable resource specifications (storage, compute, networking) tailored to
their system's specific requirements. Service providers access metal instances,
while builders customize instances for tight integration of services into ICN.
This tailored approach accommodates the specific requirements of their
integrated services.

\begin{figure}[h]
	\centering
	\includegraphics[width=1\columnwidth]{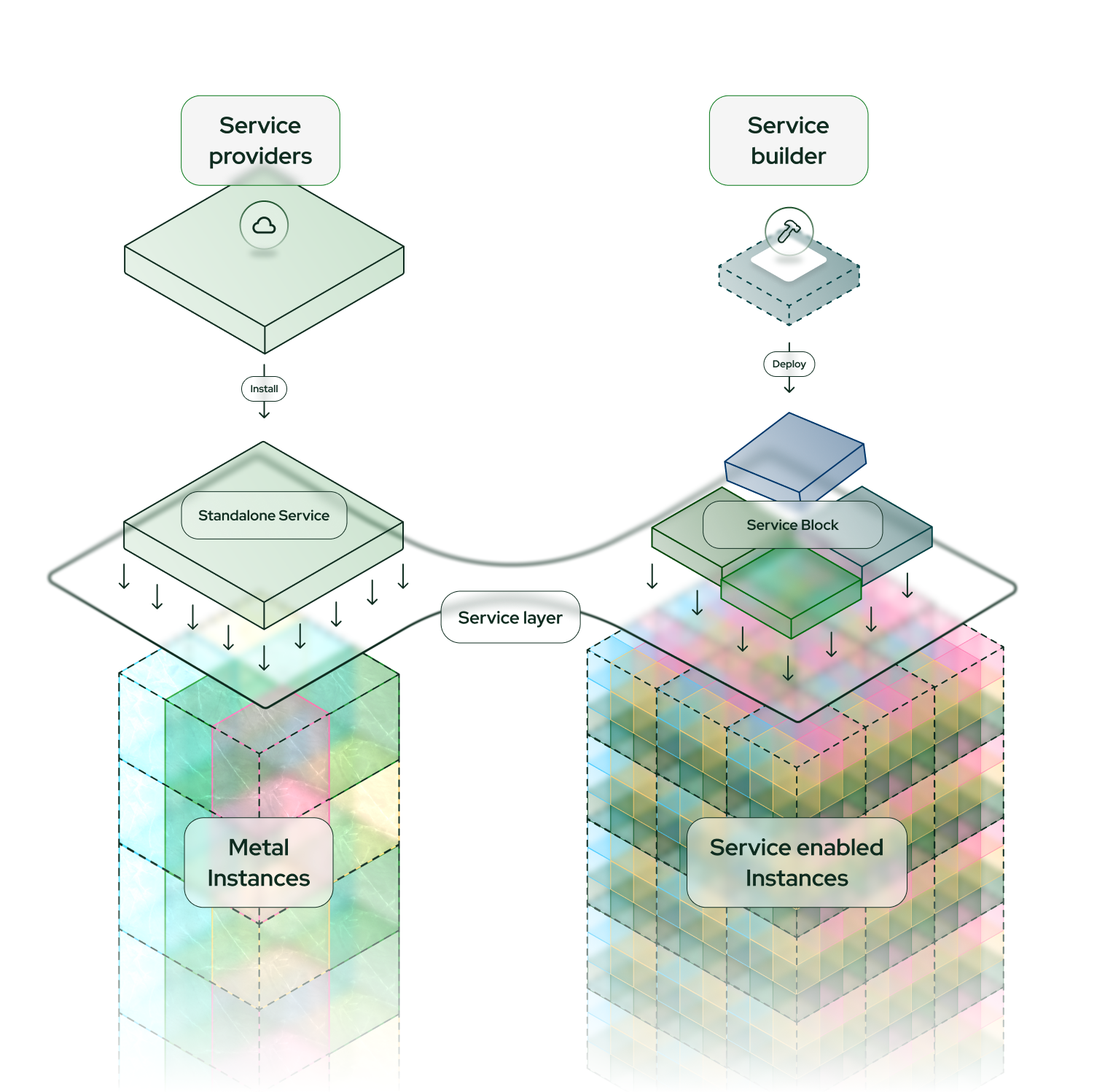}
	\caption{Service Layer}
\end{figure}

\subsubsection{Service Provider}

Service providers (SPs) deliver services by deploying software on top of ICN.
SPs reserve network capacity from ICN units where they will deploy their
services. When submitting capacity requests, this triggers the automatic
allocation of necessary resources and rewards within the ICN. Service providers
can submit capacity requests at the level of clusters where the ICN protocol
then assigns specific ScalerNodes. SPs begin payment for resources allocated by
ICN immediately.

Importantly, the SP role defined here is a borrowed structure that is typical in
traditional Web2 infrastructure. ICN intends to continue to support the existing
business models where service providers rent instances and operate their
businesses on top of it.

\paragraph{Traditional Service Providers}

Typical Service Provider (SP) setups require SPs to reserve hardware resources a
priori and pay for what they reserve. ICN remains as a hardware procurement
system similar to traditional infrastructure and we provide “metal” access to
SPs. Here the deployment of services is managed entirely by the SP, and ICN’s
involvement does not go further than resource provisioning.

In this model SPs are fully in control of their machines and deployment
processes and pay access fees for the instances they receive. All other
integrations with their deployed software are handled also by the SP and ICN
does not interfere. This model fits best the Web2-centric service where
delivering certain service solutions is a core business model.

\paragraph{Optimized deployments}

As SPs typically will back-order a significant number of machines in bulk to
fulfill their needs present and future, ICN offers optimized deployments to
allow potentially cheaper and more performant deployments. We refine resource
allocation by working with specified use cases needs, and may find hardware
configurations and software stacks that result in a more optimized end product
for the SP.

\subsubsection{Service Blocks}

Service Blocks are modular software components that are built to deploy on the
ICN Operating Subsystem. These blocks are able to be combined in a distributed
or integrated setup allowing dynamic deployment of services in elastic resource
instances or static resource instances. The composability of Service Blocks
allow the creation of macro Service Blocks that deliver a unified set of
products in a single configuration. In cases where smaller Service Blocks are
commonly used together, macro Service Blocks can ease the spin-up of
service-integrated instances.

Service Blocks are built by Service Builders which is a permissionless role that
anyone can assume. Service builders implement new services or software that are
compatible with the ICN OS resource-aware API through the SDK and integrate them
with the protocol for them to become Service Blocks. Service Blocks can range
from data storage solutions, to compute/simulation algorithms, to dedicated
servers.

Once integrated with ICN, Service Blocks can be deployed on-demand to freshly
created instances allowing a plug-and-play modality for application developers
and users alike.

\paragraph{Service Integration}
Service software built for ICN integration must be compatible with the ICN
architecture, namely the Resource Composition Layer. Instances will be created
that will be composed of various resource capabilities, and service software
must be able to support the plurality of different configurations.

Service builders will build against a clearly defined API or SDK that provides
specifications for minimal behaviour for an ICN-compliant service software. Once
it is built, it is submitted to ICN for testing and validation before being
integrated into the ICN system. Once integrated, the service will appear as an
option during instance deployment request, where a user can ask for a
service-integrated instance. This will deliver a fully capable machine with the
desired services already installed and operating.

Service integration allows users to easily spin-up tools they need for their
applications and decommission them where needed, giving deployers the power to
compose different services together for any type of product or application.

\paragraph{Automated deployment}

All services that are integrated with ICN gain the flexibility to be deployed
automatically to instances. As noted, users are able to configure their instance
and choose numerous services to pre-install into their instance. Here service
builders play no additional role in spinning up instances that operate their
service, and continue to earn rewards from its usage.

Unlike traditional service providers, service builders focus only on the
technical aspects of service delivery by building service software. They earn
rewards for their efforts when users make use of their software and they do not
have to rent resources in order to deliver value.

\section{Blockchain}

ICN Protocol is driven, in part, by smart contracts deployed to a blockchain.
The role of the blockchain in our architecture is to ensure a
censorship-resistant, transparent, irrepudiable coordination layer that will
process the core lifecycle events that determine the state of the system. We
minimize the on-chain interactions to key transactions such as value transfers,
proof submissions and access attribution to resources. In this way we leverage
the pre-existing properties of the blockchain to avoid double spend problems
related to both resource payments and reward emission, while keeping a clear
ledger of regional and global capacity of the ICN network as a whole. It becomes
a synchronization layer for the wider infrastructure to coordinate.

ICN remains blockchain-agnostic, and aims to be multi-chain in order to broaden
access to the ICN Protocol and maintain independence from specific technologies
or ecosystems. Any implementation choices to bootstrap the system on a
particular blockchain are not definitive design decisions influenced by the
protocol requirements.

\subsection{Regional Economics}

The global resource map is split into discrete regions. These regions are
similar in concept to those used by traditional cloud providers, however ICN
defines its own set of region boundaries that closely follow the desired
economic behaviour of the system. To promote a dynamic yet healthy,
self-sustaining economy, the ICN defines sets of regions that have specific
reward rates to incentivize resource provisioning by hardware providers in each
region. These incentives are designed closely with localized economies of
operating hardware to our required performance levels in mind.

In essence, the infrastructure economics of operating hardware from data centers
or other facilities must be reflected in the payout mechanism while also
allowing opportunities for hardware providers to capture value and be rewarded
for providing resources to ICN. The design of our regional system has also
maintained a heavy emphasis on the open market being a key influence on a
healthy economy; namely that demand and supply for ICN resources must remain a
driving factor for incentivizing participation. Users must be able to pay an
appropriate rate for resources and hardware providers must be appropriately
rewarded for providing them.

While there is not a strict rule for region definition, a general rule for
defining region boundaries will follow roughly homogeneous zones from the
perspective of operational cost. A region can roughly be defined as a singular
economic zone that has the same operational cost of infrastructure per unit of
resource. If a region is particularly large, it may be split into several even
if the economic zone seems similar in cases where there is a clear difference in
service accessibility or delivery e.g. latency or responsiveness gradients
across parts of the region exceed acceptable performance thresholds.

Regions will share incentive schemes and will have target capacities for each
resource. These are set up such that the region will encourage hardware
provision in high demand regions and less so in lower demand regions to reflect
market dynamics. Each region will have a bootstrap period where incentives are
paid out to early participants to each region. After the bootstrap phase is
over, incentives will shift to access fee payments for resources to continue
rewarding hardware providers.

\subsection{Hardware Collateralization}

Within the ICN, hardware collateralization is a fundamental requirement,
securing the network and guaranteeing hardware reliability by holding hardware
providers accountable for their performance. Hardware providers lock tokens as
collateral, in proportion to the hardware resources they provide, which remain
locked for the entire commitment period, demonstrating their dedication to
meeting network obligations. This collateral serves two main purposes: ensuring
network reliability through penalizing underperforming hardware providers via
slashing, and encouraging sustained participation to enhance network stability.

\subsection{Ecosystem Token}

The ICN Protocol operates on a core asset for network balance. It is the primary
vehicle for collateral in the ICN Protocol, where it is used by users and
hardware providers both to activate and secure hardware resources. The Hardware
Layer relies heavily on the token to form a robust resource network and
incentivize good performance and provisioning while punishing bad behaviour and
faults.

This ecosystem token is the sole collateral asset for hardware providers to
achieve threshold collateral required to operate. Each ScalerNode has a minimum
node collateral requirement based on the resources provided and must be met in
order for the resource to be active. Once active non-hardware providers are able
to also stake tokens to these resources, which acts as an additional layer of
security for the hardware network. Misbehaving ScalerNodes will have their
collateral slashed proportionally to the severity of the fault.

The token also has a utility in the ICN system in accessing resources in the
hardware network. These take the form of access fees that are paid in the token
for the usage of resources by users or builders.

\subsection{NFT Pass}

Within the ICN Protocol exists a secondary asset that serves as a stakeable NFT
module. It functions as a bootstrapping mechanism that powers the genesis of the
network with pre-committed security. By staking the NFT to either the
performance enforcement layer or the hardware network layer (in a 1:1
NFT-to-Node relationship), users contribute security to the network. Each NFT
upon creation includes a time-locked value sink that diminishes over time. Upon
staking, this sink value acts as collateral for the designated node, and its
decay transforms into a continuous token reward for the staker. Similar to the
token, the staked NFT is susceptible to slashing, which accelerates its decay
upon node faults or misbehavior. Once the time-lock sink fully decays, the NFT
no longer provides security to the network. NFT owners are incentivized to stake
to benefit from the decaying value as a reward, concurrently providing the
remaining value sink as accumulated security collateral.

\subsection{Proof Verification}

Proof verification within the ICN protocol is integral to maintaining the
integrity and reliability of its off-chain operations. The HyperNode Network, as
a validator network, generates performance assessments of the hardware layer.
These assessments, along with key performance indicators, are published
regularly to the Satellite Network, which ensures data availability and
auditability. Crucially, cryptographic proofs derived from these assessments are
submitted to the blockchain. This on-chain proof submission allows for
transparent and public verification that the reports committed to the Satellite
Network accurately align with the validated performance data, thereby fostering
trust and accountability across the entire network.

\subsection{Resource Reservation}

The ICN protocol streamlines the allocation of network resources by
automatically fulfilling requests based on optimal performance, pricing, and
availability within a given region. This automated selection process means that
users cannot directly choose specific hardware configurations.

For users requiring guaranteed resources, advance reservations are possible for
a defined duration. While extensions to these reservations are possible, they
are contingent on the hardware provider accepting the associated increased price
risk. It's important to note that the access fee for the token is fixed at the
time of booking, which effectively transfers the price risk for both extended
and future resource requests to the ICN protocol and the hardware provider that
ultimately fulfills the request.

\section{Conclusions}

The cloud landscape has been dominated by a handful of centralized hyperscalers
that dictate the terms of innovation, data sovereignty, and access. This
consolidation, while offering convenience, has created a fragile ecosystem prone
to censorship, single points of failure, and a fundamental disconnect between
users and their data. The Impossible Cloud Network emerges as a direct response
to these challenges, presenting a new paradigm for a truly decentralized
internet infrastructure. 

By decoupling the hardware, resource composition, performance enforcement, and
service layers, ICN creates a resilient, transparent, and permissionless
ecosystem. The protocol's multi-layered architecture, governed by a
blockchain-agnostic coordination layer, ensures that no single entity can exert
undue control. The novel Performance Enforcement layer, with its network of
HyperNodes, introduces a level of verifiable trust and accountability previously
unseen in decentralized systems, guaranteeing enterprise-grade performance.

The economic model, centered around a native token and hardware
collateralization, creates a self-sustaining and robust economy that
incentivizes participation and good behavior from hardware providers across the
globe. This promotes a diverse and competitive environment for resources,
breaking the vendor lock-in that characterizes the current cloud industry.

Future work will focus on expanding the ICN ecosystem. Key next steps include
growing the global network of hardware providers, incentivizing the development
of a rich library of services, and refining the protocol's governance mechanisms
to ensure long-term decentralization and community ownership. We will also focus
on building out developer tooling and documentation to lower the barrier for
entry and encourage widespread adoption.

\begin{appendices}

\section{Use Cases}

\subsection{Porting Web2 classic cloud to Web3}

More than half of global IT infrastructure spending is directed towards public
cloud services. Industry analyst Gartner estimates this \$723.4 billion market
will continue to grow at a rate of 21.5\% year over year
\cite{gartner_2024_gartner}. The public cloud infrastructure market is dominated
by only a handful of centralized “hyperscaler” platforms, such as Amazon Web
Services (AWS), Microsoft Azure and Google Cloud Platform (GCP) in the West, and
in China,  Alibaba, and Tencent. This oligopolistic environment has caused
vendor lock-in and inflated pricing \cite{10.1109/MIC.2013.19, 6682808}, suffers
from single-points-of-failure causing wide-scale outages
\cite{pendleton_2024_cloud}  and raises significant security, privacy and
ownership concerns around user vs. corporate/government control
\cite{doi:10.1177/2053951720982012}. 

ICN's web3 architecture is fully compatible with conventional workloads, as the
majority of this global compute, AI and storage demand is driven by Web2-style
applications, including centralized software-as-aservice (SaaS) platforms,
enterprise backends, data and business processing workflows, and other services.
ICN allows seamless portability of these existing business operations as it
offers a decentralized, hardware-backed alternative that enables the deployment
of compute, storage, and networking environments through composable instance
blueprints. These environments are functionally equivalent to virtual machines,
Kubernetes clusters, or bare metal servers, ensuring portability of existing use
cases, and can be configured to meet specific workload requirements.
Additionally, ICN addresses the concerns mentioned in the previous paragraph
through its web3 features like transparency, verifiability and sovereignity.

ICN supports two deployment models. In the Metal-as-a-Service approach,
enterprise operators can reserve physical or virtualized compute instances for
fixed durations. This enables deterministic pricing and alignment with
service-level or compliance requirements, and is suited to use cases requiring
fine-grained control over hardware allocation and lifecycle. Alternatively,
Elastic, On-Demand Instances allow DevOps teams to provision compute resources
for applications with dynamic or auto-scaling requirements. These instances can
be launched using predefined blueprints or customized configurations via the
Resource Composition Layer. The platform supports vertical and horizontal
scaling, lifecycle management via programmable APIs and regional deployment
constraints allowing users to meet GDPR, SOC2 and other compliance requirements
\cite{europeanunion_2016_regulation, a2023_2017}. Both of these deployment
models are compatible with existing deployment tools and architectures. They
offer usage-based pricing with transparent cost structures, global distribution
through a decentralized network of hardware providers, and network-verified
performance and availability. This approach eliminates reliance on centralized
cloud vendors, incrementally adopting  a decentralized infrastructure model,
while allowing operators to maintain existing workflows, ensuring business
continuity.

\subsection{Decentralization as the Next Frontier in AI}

The evolution of Artificial Intelligence is signaling a strategic shift from
centralized architectures to a more distributed, decentralized cloud
infrastructure. Historically, the initial surge in AI development has been
largely tethered to centralized hyperscaler clouds. These platforms, while
enabling significant advancements in model training, are increasingly facing
architectural limitations in the face of emerging AI paradigms. The move towards
decentralized cloud infrastructure is driven by several key factors:

\paragraph{Evolving Model Training Paradigms} Recent innovations, exemplified by
	models like Deepseek, demonstrate a diminishing dependency on massive
	centralized training clusters. Deepseek's architectural efficiencies,
	including optimized attention mechanisms (e.g., Multi-Head Latent Attention)
	and sparse Mixture-of-Experts (MoE) approaches, allow for competitive
	performance with significantly reduced computational resources
	\cite{deepseekai2025deepseekv3technicalreport, moore_2025_the}. A bright
	roadmap still lies ahead when this will be combined with techniques from
	Differential Privacy \cite{10.1145/2976749.2978318} and Federated Learning
	\cite{MAL-083}. These advancements indicate a progression towards more
	efficient and distributed training methodologies, democratizing AI
	development and making it accessible to a broader range of participants
	beyond well-capitalized entities. This shift inherently favors decentralized
	compute networks, which can aggregate and orchestrate global computational
	resources.

\paragraph{Edge Inference} The necessity for AI inference to occur at the edge,
	closer to the end-users and data sources, is becoming paramount.
	Applications such as autonomous systems, real-time augmented reality, and
	localized IoT analytics demand minimal latency for effective operation.
	Centralized inference introduces inherent delays due to data transmission
	distances. Decentralized cloud networks, by distributing computing power
	geographically, facilitate low-latency inference at the network's periphery.
	This architectural advantage is critical for enhancing user experience and
	enabling new categories of real-time AI applications especially when
	combined with the critical need for confidentiality and verifiability
	achieved through techniques like Zero-Knowledge Machine Learning (ZKML)
	\cite{zhang2024secureprivateaiframework}.

\paragraph{The Rise of Agentic AI}  The forthcoming AI cycle is anticipated to
	be dominated by agentic AI, characterized by autonomous entities capable of
	complex decision-making and interaction. These agents, whether coordinating
	in swarms or operating independently, require not only continuous operation
	and distributed coordination but also exceptionally low latency for their
	localized data processing and real-time decision-making. This critical need
	for minimal delay stems from the iterative nature of agentic workflows:
	agents often operate on a sense-think-act loop
	\cite{krishnan2025aiagentsevolutionarchitecture}, where even minor delays in
	sensing new information or executing an action can lead to suboptimal
	outcomes, missed opportunities, or even critical failures in dynamic
	environments.This makes them inherently better suited for a decentralized
	cloud environment. Centralized systems introduce inherent latency due to
	data round-trips to distant servers, alongside presenting single points of
	failure, potential censorship vectors, and economic bottlenecks. These
	factors could severely impede the ubiquitous and resilient deployment of AI
	agents. A decentralized cloud, with its distributed compute resources closer
	to the edge, provides the essential resilience, autonomy, scalability, and
	crucially, the low latency vital for the robust functioning of agentic
	systems.

\subsection{NaaS providers}
Node-as-a-Service (NaaS) providers have emerged as a crucial utility in the Web3
ecosystem, simplifying access to blockchain networks for developers. By
abstracting away the complexities of running and maintaining full nodes, they
have significantly lowered the barrier to entry for running decentralized
applications. Through pre-configured, deployable nodes on the cloud, they
eliminate the need for both technical operational expertise as well as hardware
resources, allowing developers to focus on building innovative DApps and smart
contracts.

However, a critical conflict persists: the vast majority of these NaaS providers
currently operate on centralized hyperscaler cloud infrastructure, like Amazon
AWS, Microsoft Azure, Google GCS or Hetzner, or run their own single datacenter.
The vulnerability of this reliance becomes apparent when major parts of web3 are
unavailable when these systems encounter downtime, which even for the
hyperscalers has kept on happening. This reliance on centralized systems
fundamentally contradicts the core tenets of Web3 - decentralization, censorship
resistance, user ownership and no single points of failure. 

ICN is building the foundational layer for the next generation of the internet,
explicitly aiming to provide the infrastructure housing the base-layer Web3
projects. This means providing the underlying decentralized compute, storage,
and networking resources that are essential for blockchains and decentralized
applications to operate with true Web3 integrity.

The evolution of Web3 demands that its supporting infrastructure reflects its
core values. The current state of NaaS, while convenient, represents a
significant centralizing force within an ecosystem explicitly designed to be
decentralized. NaaS providers have a critical role to play in this transition by
moving away from centralized cloud giants and embracing the decentralized
future. This shift is not just about technological advancement; it's about
upholding the fundamental principles that define the Web3 revolution.

\end{appendices}

\printbibliography

\end{document}